\documentclass[12pt]{iopart}
\usepackage{graphicx}
\usepackage{cite}
\usepackage{epsfig}
\usepackage{subfigure}
\usepackage{wrapfig}

\begin{document}

\title[Heavy-Flavor
  Measurements by the PHENIX Experiment at RHIC]{Heavy-Flavor
  Measurements by the PHENIX Experiment at RHIC}

\author{Marisilvia Donadelli for the PHENIX Collaboration}

\address{Universidade de Sao Paulo, Instituto de Fisica, Caixa Postal
  66318, Sao Paulo CEP 05315-970, Brazil}
\ead{mari@rcf.rhic.bnl.gov}
\begin{abstract}

In recent years, PHENIX has studied many important observables related
to heavy-flavor physics through their leptonic decay measurements
including the invariant yield of electrons from nonphotonic sources,
and prompt single muons, both of which are dominated by D and B
mesons.  Charm and beauty cross-sections were  measured and compared 
through single lepton, and lepton-hadron correlations  in $p$+$p$
collisions at $\sqrt{s}$ = 200 GeV.  Observables for quarkonia production such as
invariant yield and polarization were also measured in $p$+$p$
collisions. In Au+Au collisions, preliminary results for the $R_{AA}$ for single
electrons and a 90\% CL upper limit for the suppression of
$\Upsilon$s were produced. And in $d$+Au collisions, a preliminary
$R_{CP}$ study for $J/\psi$ production in different centrality ranges
was extracted.

\end{abstract}

\submitto{\JPG}

\section{Introduction}
Measurements of heavy-flavor production in $p$+$p$
interactions at collider energies serve as important tests for
perturbative quantum chromodynamics (pQCD). It also provides an
important baseline for study of the medium created in relativistic
heavy ion collisions, since the transition from hadronic matter to a
quark gluon plasma at high temperature is expected to be achieved in
such collisions. On the other hand, the hadronization of partons in vacuum or cold
nuclear matter, represents a non perturbative QCD process. In order to measure
the baseline, disentangle the cold nuclear matter effects and
determine the hot nuclear matter effects, the  Relativistic Heavy Ion
Collider (RHIC) provides, respectively, the 
 colliding species: $p$+$p$, $d$+Au and Au+Au or Cu+Cu. 

The PHENIX \cite{Adcox2003469} experiment at RHIC has 
capabilities to measure both the leptonic and hadronic decay modes of
the particles produced in the collisions. That allows a
comprehensive study of open heavy-flavor and
heavy quarkonia.  Electrons are identified in two central arms
spectrometers covering the pseudorapidity range $|\eta| <$ 0.35 and 2
$\times \pi / $2   in azimuth.  Drift chambers and pad chambers are
used for charged particle tracking, a ring imaging  $\breve{C}erenkov$
detector for electron identification and an electromagnetic
calorimeter for electron identification and trigger. Muons are
measured in two separate spectrometers with full azimuthal coverage
and in the pseudorapidity range 1.2$ \leq |\eta| \leq $ 2.4.  They are
identified in panels formed by Iarocci tubes placed in front of steel
plates for hadron absorption. While
the muon arm absorber removes most of the hadrons, they are also
measured in the central arms.  In addition, beam-beam counters
positioned at pseudorapidity 3.1 $ \leq |\eta| \leq$ 3.9, measure the
position of the collision vertex along the beam direction and provide
the interaction trigger. In A+A collisions, they also
provide the event centrality characterization through a correlation of
the beam-beam counters charge sum and the total energy deposited in
the two zero degree hadronic calorimeters, located at $\sim$ 18 m from
the interaction point. 
 
\section{Open heavy-flavor}

In PHENIX, since D or B mesons are not fully reconstructed, their yield
can be statiscally derived by their semileptonic decays contributing
to lepton spectra.  These indirect  measurements serve as baseline for both charm
and bottom production in  $d$+Au  and A+A collisions.

\subsection{{\bf Single leptons}} 
\begin{wrapfigure}{r}{0.49\textwidth}
{\includegraphics[width=0.49\textwidth]{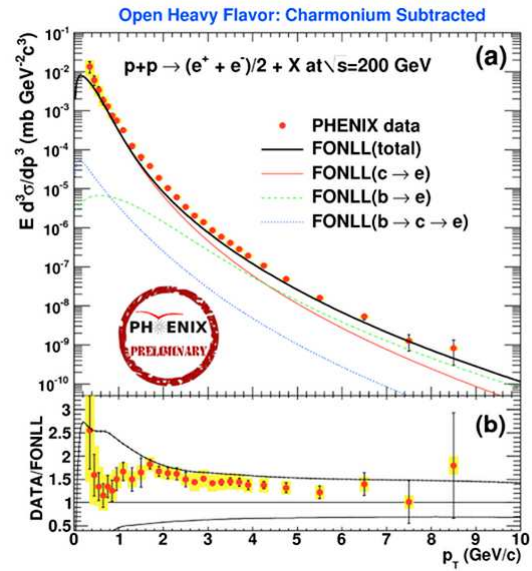}}
\vspace{-15pt}
  \caption{(a) Invariant cross  sections of
    electrons after subtraction of the $J/\psi, \Upsilon$,
    Drell Yan contributions \cite{dion-2009}. (b) Ratio of the data
    and the FONLL calculation. The upper (lower) curve shows the
    theoretical upper (lower) limit of the FONLL calculation.}
  \label{fig:espectrum}
\end{wrapfigure}

The spectrum of
electrons from heavy-flavor decay in $p$+$p$  collisions was revisited
\cite{dion-2009}
as an additional contribution to the previous spectrum of electrons
\cite{adare:252002}. The inclusive electron spectra consist of
nonphotonic electrons from heavy-flavor decays; photonic background
from Dalitz decays and photon conversions in the beam pipe; 
nonphotonic background from $K_{e3}$ and dielectron decays of vector
mesons. The  $K_{e3}$ contribution is small compared to the photonic
background from Dalitz decays. The electron spectrum from heavy-flavor decay was determined by subtracting the
background components from the inclusive spectrum using two
independent and complementary methods.  The electron
spectrum from all known sources  except semileptonic decay of heavy-flavor is calculated using a Monte Carlo simulation and subtracted
from the inclusive spectrum in the cocktail method.  The dominant source
of background is the $\pi^{0}$ Dalitz decay, so the measured $\pi^{0}$
and $\pi^{\pm}$ spectra are used as input to the generator. The cocktail also
includes contribution from quarkonium ($J/\psi$, $\Upsilon$) and the
Drell-Yan process, which were neglected in
\cite{adare:252002,adare:172301} . Figure \ref{fig:espectrum} shows
the nonphotonic electron yield after the subtraction of the 
$J/\psi$ (that accounts for up to 16\% of the electron signal at 
$p_T >$ 5 GeV/c), the $\Upsilon$ and the Drell-Yan contributions (both found to be
negligible) in $p$+$p$  collisions \cite{dion-2009} .
The bottom part of the figure shows the ratio of data to FONLL
calculations \cite{PhysRevLett.95.122001}.  From the
nonphotonic electron spectrum derived in \cite{adare:252002}  the total charm production
cross-section was deduced to be $\sigma_{c\bar{c}}$ = 567
$\pm$57 (stat) $\pm$ 193 (sys) $\mu b$.   

Similar measurements were performed for single muons in which a
background cocktail approach is also used. Backgrounds in the muon arms are divided
into muons that result from the weak decay of light hadrons before
the first absorber material and hadrons that penetrate the steel
absorber to reach the deepest layer in the muon arm. Muons from
heavy-flavor meson decay originate $<$ 1mm from the collision vertex,
while yields of muons from hadron decay exhibit a linear collision
vertex dependence before the first absorber material due to decay
kinematics.  The background
is then subtracted using vertex distribution and hadrons which are stopped
in the muon identifier.

Heavy-flavor single muon measurements in $p$+$p$
collisions at $\sqrt{s}$ = 200 GeV and for $\langle$ y $\rangle$ =  1.65 
were reported in \cite{adler:092002},  

\begin{wrapfigure}{r}{0.49\textwidth}
{\includegraphics[width=0.44\textwidth]{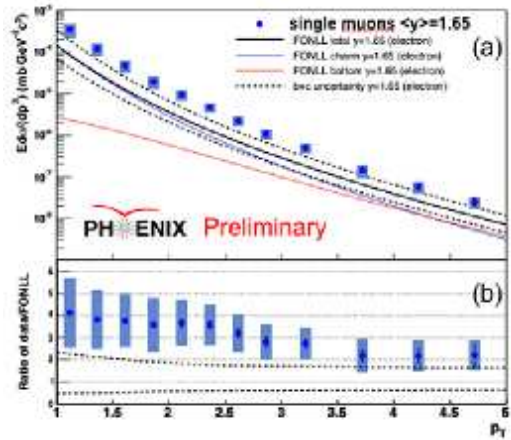}}
\vspace{-8pt}
  \caption{Invariant cross sections of muons
    from heavy-flavor decay at $\langle y \rangle$ = 1.65
    \cite{0954-3899-35-10-104113}. (b) Ratio of the data
    and the FONLL calculation. The upper (lower) curve shows the
    theoretical upper (lower) limit of the FONLL calculation. }
  \label{fig:mspectrum}
\end{wrapfigure} 

\noindent with a corresponding diferential
cross section for charm quark production at forward rapidity of $d
\sigma_{c\bar{c}}/dy|_{y=.6}$ = 0.243 $\pm$ 0.013 (stat) $\pm$ 0.105
(data syst.) $^{+0.049}_{0.087}$ (PYTHIA syst.) mb. The new result at $\langle$ y $\rangle$ =  1.65 
was obtained over the extended range of 1 $\leq p_T \leq$ 5.0 GeV/c
\cite{0954-3899-35-10-104113}. Single muons are measured in both
forward and backward directions, with the two spectra being combined
to form the spectrum shown in Figure \ref{fig:mspectrum}
\cite{0954-3899-35-10-104113}, which is compared to a FONLL
calculation over the same rapidity.

In Au+Au collisions, the heavy-flavor signal and the ratio
of nonphotonic to photonic electrons were determined via exactly the
same independent and complementary methods  used in  \cite{adare:252002}. Similarly to the $p$+$p$ case, the contribution
of the $J/\psi$ to the electron spectrum in Au+Au collisions was
determined \cite{dion-2009}.  Two conservative estimations of the upper and the lower
bound of the $J/\psi$ yield at high $p_T$ were made using the measured
nuclear modification factor
$R_{AA}$ \cite{adare:232301} in the 0-20 \%  centrality range.  

\begin{wrapfigure}{r}{0.49\textwidth}
    {\includegraphics[width=0.49\textwidth]{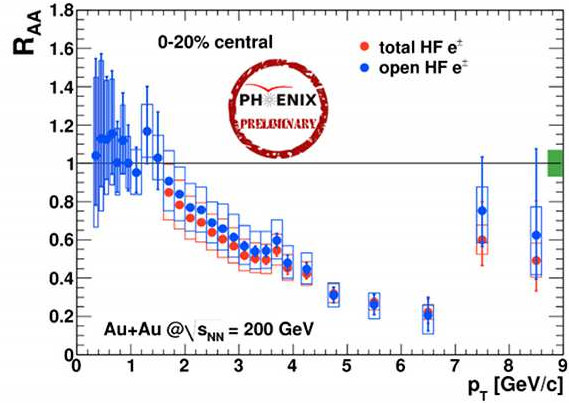}}
\label{fig:espectrumauau}
\vspace{-25pt}
  \caption{$R_{AA}$ of nonphotonic
    electrons with and without $J/\psi$  contribution
    \cite{dion-2009}. }
\end{wrapfigure}

\noindent  PHENIX has
previously presented the $R_{AA}$ result in the 0-10 \% centrality
range  \cite{adare:172301} showing a suppression level for the
heavy-flavor electrons almost the 
same as for $\pi^{0}$ and for $\eta$ in high $p_T$ region.  For the 0-20 \% centrality range, as that how $J/\psi$  was
measured, the result with
and without its  contribution can be seen in Figure
3. Both  the  $p$+$p$
and the Au+Au single electron yields from heavy-flavor decays
decreased after subtracting the  $J/\psi$ contribution, resulting
 on a small change in the  $R_{AA}$.

\subsection{{\bf Diparticle correlations}} 

\begin{wrapfigure}{r}{0.49\textwidth}
\vspace{-20pt}
    {\includegraphics[width=0.52\textwidth]{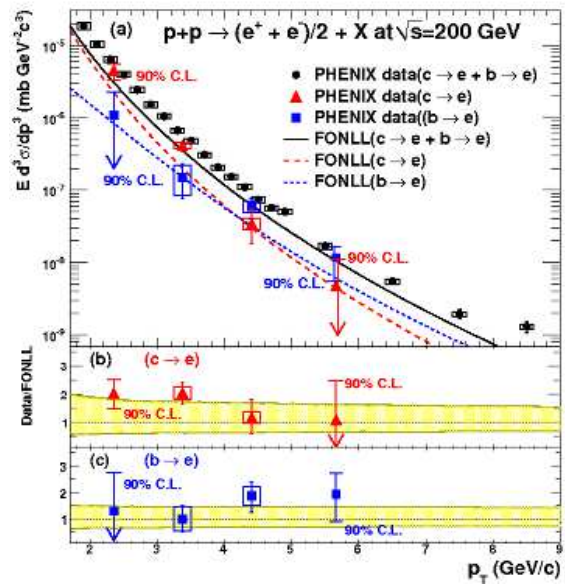}}
\label{fig:eh1}
\vspace{-25pt}
  \caption{Invariant cross sections of electrons from charm and bottom
    \cite{adare:082002} with the FONLL calculation \cite{PhysRevLett.95.122001}. }
\end{wrapfigure} 

In PHENIX,  D meson decays cannot be fully reconstructed because of
the small signal/background ratio and the poor PID of kaons. The ratio of
$(b \rightarrow e)$ to $ [ (c \rightarrow e)+ (b \rightarrow e)]$
as a function of electron $p_T$  in the range 2 $< p_T < $ 7 GeV/c was
extracted from the correlation between the heavy-flavor electrons and
associated hadrons and was used to separate the beauty from the charm contribution
\cite{adare:082002}. The extraction was based on partial reconstruction of
the $D/\bar{D} \rightarrow e^{\pm} K^{\pm} X $ decay. The
electron-hadron  invariant mass is formed and exhibits a clear  peak below the $D$
mass.  The shape and the efficiency of this peak depend on the charm or
the beauty origin of the electron-hadron pair. The beauty decays
present a lower efficiency and broader shape, because of their mass
and alowed range of hadrons. 
In Figure 4,
the single electron spectra for charm and bottom were measured from the
ratio  $(b \rightarrow e) / [ (c \rightarrow e)+ (b \rightarrow
e)]$, and the spectrum from heavy-flavor decays.

Electron-muon azimuthal correlations in $p$+$p$
collisions have proved to be a new method for the $c\bar{c}$
measurement at PHENIX \cite{engelmore-2009-830}. The heavy-flavor signal consists of opposite sign
electron-muon pairs with $\Delta \phi \approx \pi$, product of
back-to-back charm pairs produced mainly via gluon fusion. 
 
\begin{wrapfigure}{r}{0.49\textwidth}
  \begin{center}
    {\includegraphics[width=0.49\textwidth]{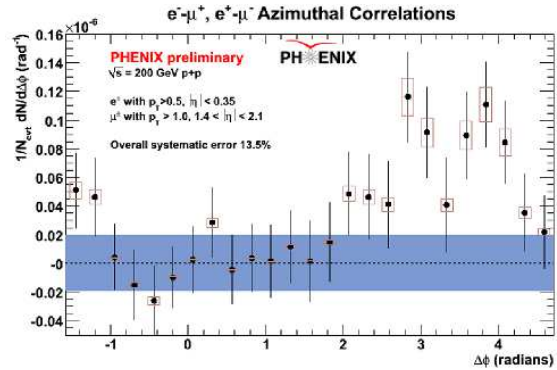}}
 \end{center}
\label{fig:eh2}
\vspace{-20pt}
  \caption{Electron-muon correlations from heavy-flavor \cite{engelmore-2009-830}.}
\end{wrapfigure}    
 
There is no contribution
from Drell-Yan, thermal production, resonance decays for the
background. Like-sign subtraction removes combinatorial background and electron-muon
pairs from dijets.  Substantial backgrounds in analysis are due to
misidentified muons: hadrons that punch through absorbers and muons
from light meson decays rather than from heavy-flavor. The background from hadron decays into muons was estimated and removed from the $\Delta \phi$ correlation distribution. Figure
5 shows the background subtracted $\Delta \phi$
distribution. The band is error from decay muon subtraction, boxes are
errors from  electron-hadron subtraction. This new method provided an
invariant yield of 2.11 $\times$ 10$^{-7} \pm$ 3.4 $\times$ 10$^{-8}$
(stat.) $\pm$ 3.5 $\times$ 10$^{-8}$  (sys.) and is a proof of principle for future charm
cross section measurement in intermediate rapidity.

\section{Quarkonia}

\begin{figure}[h]
\centering
\vspace{-30pt}
\subfigure[]{\includegraphics[width=0.45\textwidth]{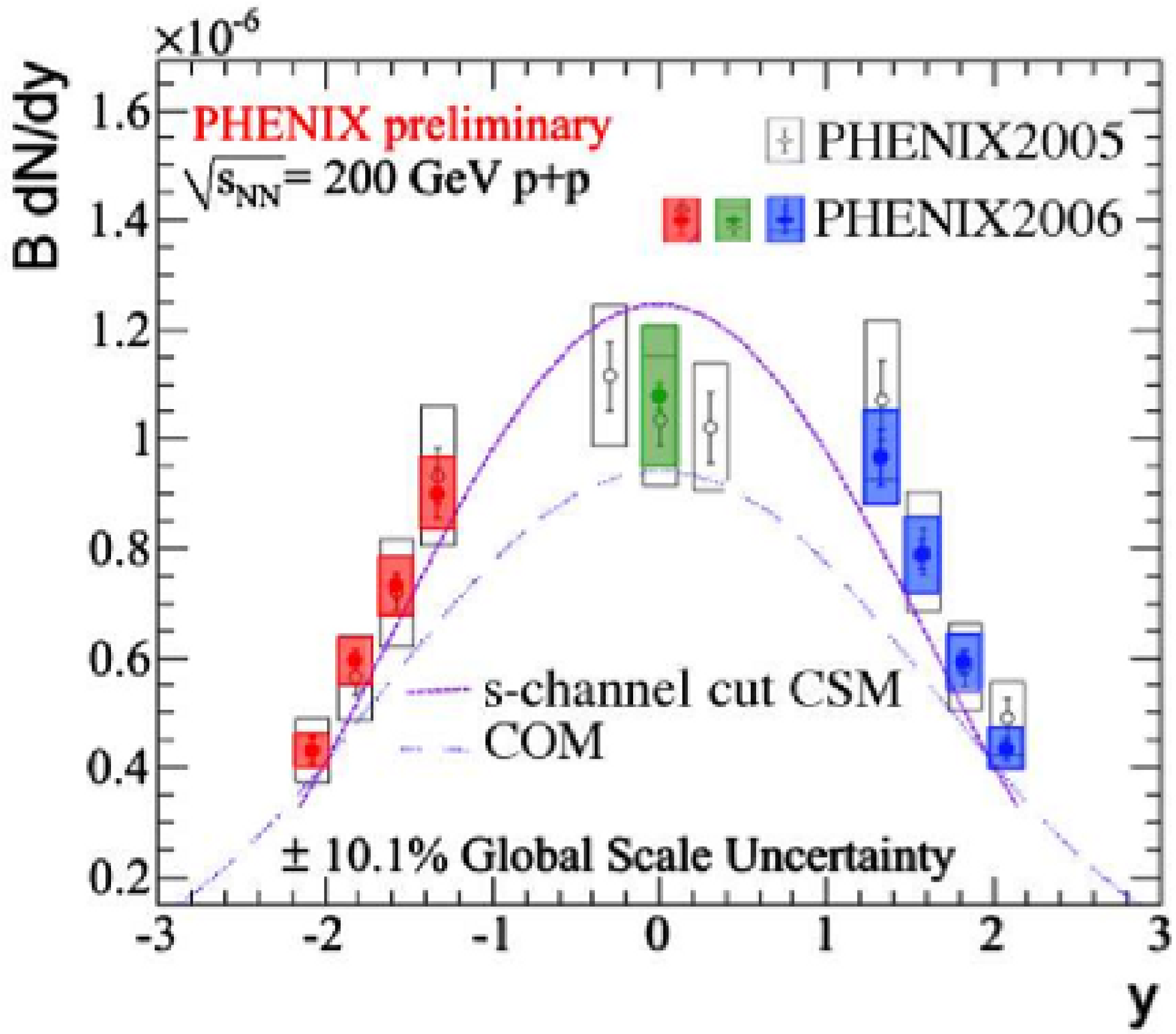}}
\subfigure[]{\includegraphics[width=0.35\textwidth]{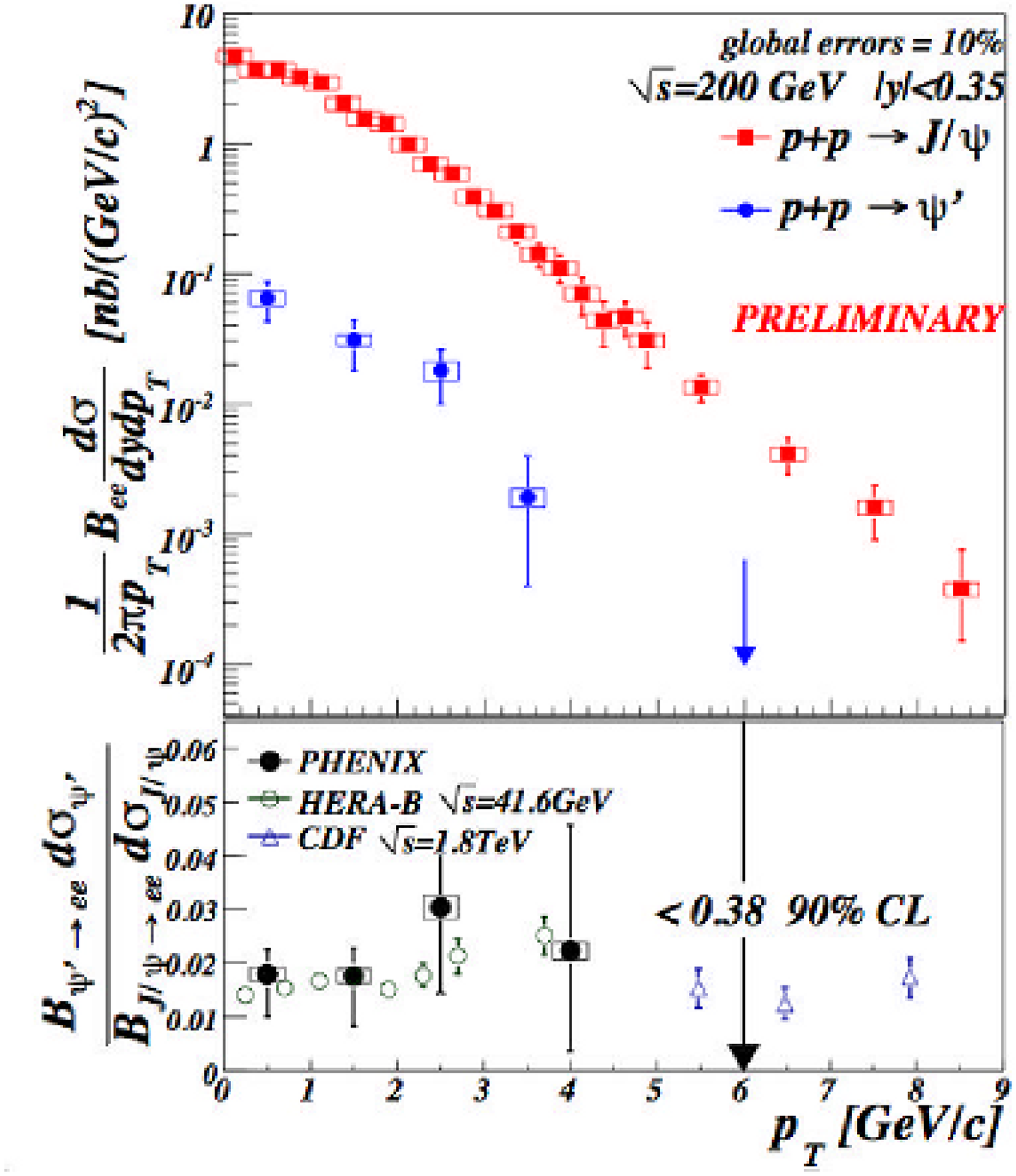}}
\vspace{-15pt}
  \caption{(a) Preliminary measurement of $J/\psi$ yield multiplied by
    the branching ratio versus rapidity with both COM \cite
    {PhysRevLett.93.171801} and s-channel cut Color Singlet Model
    (CSM)\cite{haberzettl:032006} calculations. (b) $p_T$ dependence
    of $J/\psi$ and $\psi^{\prime}$ at midrapidity, together with the $J/\psi$ to $\psi^{\prime}$ cross sections ratio \cite{dasilva-2009-830}
    compared to that from
other experimental results \cite{refId,PhysRevLett.79.572} . }
  \label{fig:charmonium1}
\end{figure}

Using the 2006 data taking period, PHENIX has recently released a set
of measurements on $J/\psi$ production in $p$+$p$ collisions \cite{dasilva-2009-830} including new results on the $p_T$
dependence of its decay angular distribution in
different reference frames \cite{collaboration-2009}. The record of an appoximate factor of three
of integrated luminosity, compared to the
published one \cite{adare:232002}, allowed not only to extract the differential cross section of
$J/\psi$, but also the first $\psi^{\prime}$  $p_T$ dependent
result at RHIC energies \cite{dasilva-2009-830}. Figure
\ref{fig:charmonium1} (a) shows the preliminary measurement of the
$J/\psi$ invariant yield versus rapidity, where  the uncertainties are
dominated by systematics with both forward and midrapidity yields well
described by the s-channel cut Color Singlet Model (CSM)\cite{haberzettl:032006}. Figure
\ref{fig:charmonium1} (b) shows the differential cross section of both
$J/\psi$ and $\psi^{\prime}$. The $\psi^{\prime}$ to $J/\psi$ cross
sections ratio is shown in the lower pannel and compared to that from
other experimental results \cite{refId,PhysRevLett.79.572}.  The
feed-down contributions to $J/\psi$ from $\psi^{\prime}$ is 8.6 $\pm$
2.5\% and from $\chi_c$ is $<$ 42\% (90 \% C.L.) and both are in
agreement with world average results \cite{1126-6708-2008-10-004}.

\begin{figure}[h]
\centering 
\vspace{-15pt} 
\subfigure[]{\includegraphics[width=0.39\textwidth]{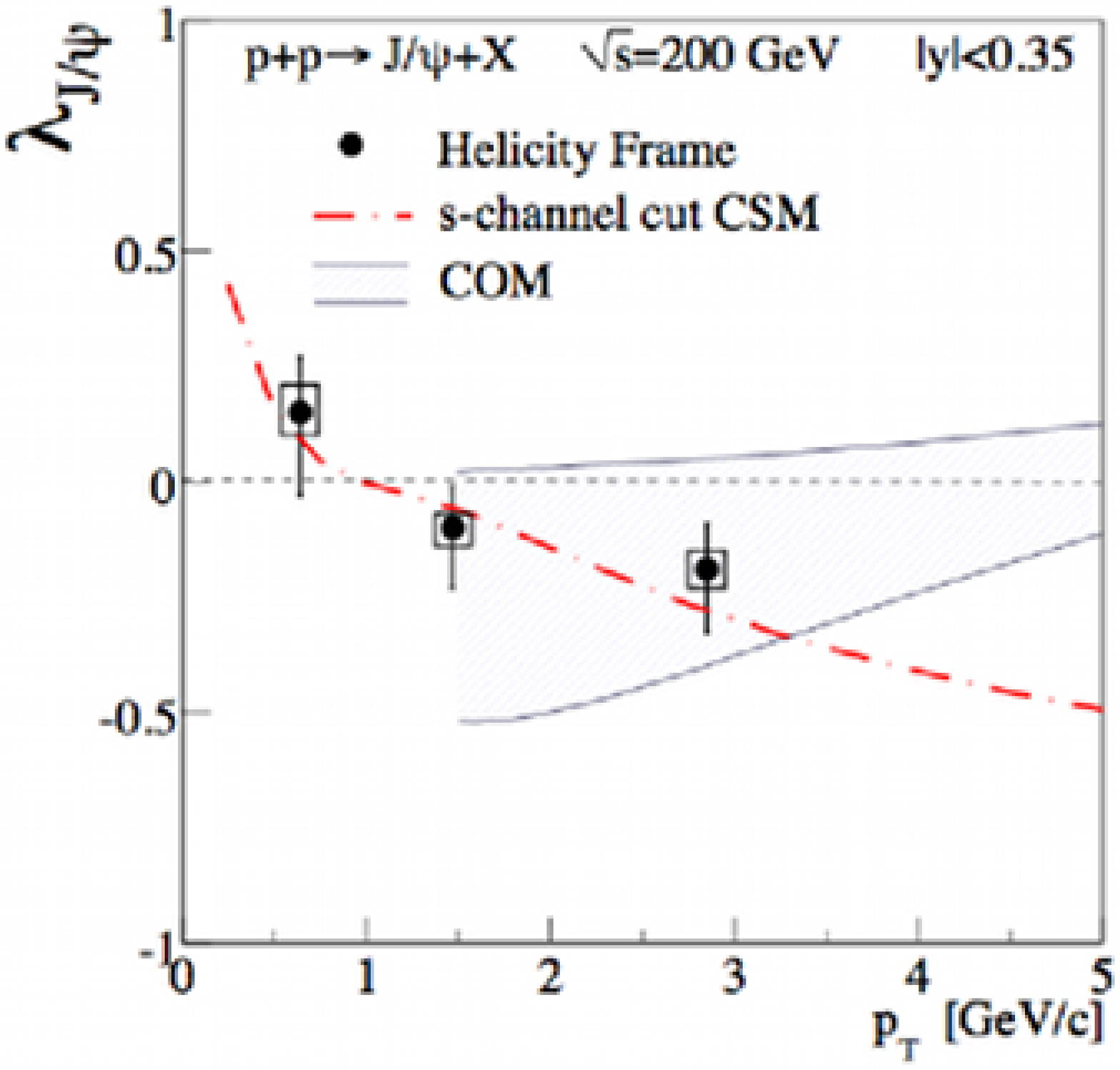}}
\subfigure[]{\includegraphics[width=0.35\textwidth]{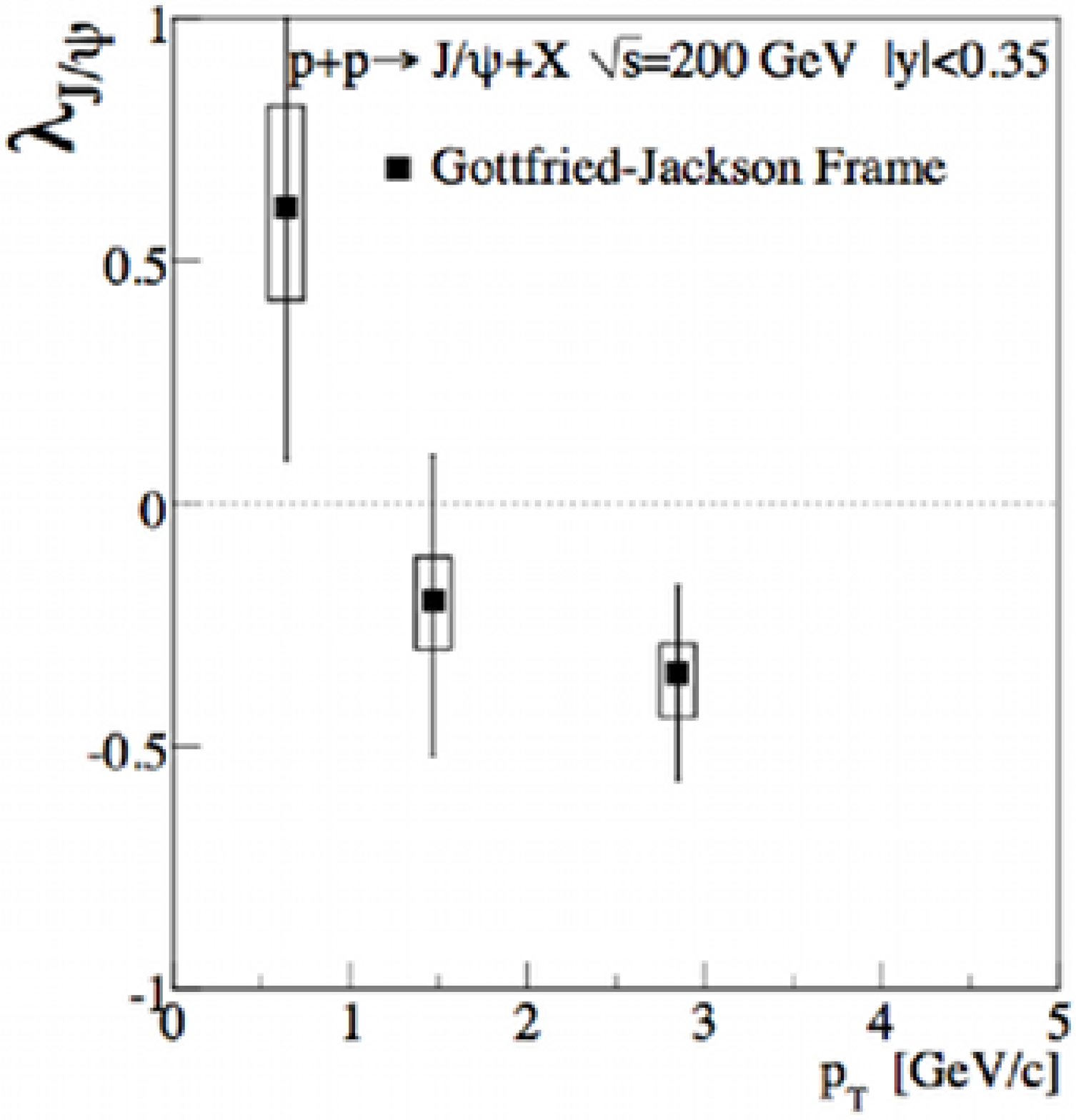}}
\vspace{-15pt}
  \caption{(a) $J/\psi$ polarization parameter versus transverse
    momentum in the HX frame and theoretical calculations. Boxes are
    correlated point-to-point
    systematic uncertainties. (b) $J/\psi$ polarization parameter versus transverse
    momentum in the GJ frame. Boxes are correlated point-to-point
    systematic uncertainties\cite{collaboration-2009}. }
  \label{fig:charmonium2}
\end{figure} 

With the same improved data set, the measurement of the transverse
momentum dependence of the inclusive $J/\psi$ polarization can help
elucidate the production mechanisms \cite{collaboration-2009}. Moreover, it is expected that the
polarization of  $J/\psi$ is modified by the presence of nuclear
matter effects in $d$+Au  collisions and hot and dense matter in Au+Au
collisions \cite{PhysRevC.68.061902}.  The polarization was studied in three reference frames: helicity (HX),
Gottfried-Jackson (GJ)  and
Collins-Soper (CS) for $p_T< $ 5 GeV/c
and $|y|< $0.35.  

\begin{wrapfigure}{r}{0.49\textwidth}
  \begin{center}
\vspace{-30pt}
    {\includegraphics[width=0.39\textwidth]{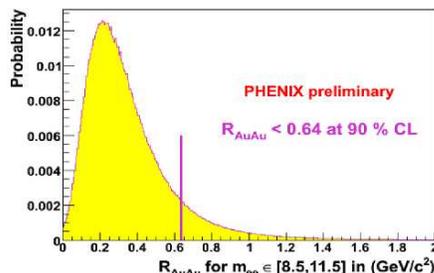}}
 \end{center}
\label{fig:upsi}
\vspace{-20pt}
  \caption{Probability distribution of
    $R_{AA}$ in the $\Upsilon(1S+2S+3S)$ mass range.}
\end{wrapfigure}  

\noindent Both results for the polarization parameter in the
HX and GJ frames are consistent with zero for the
full $p_T$ range and with a 1.8 sigma trend towards longitudinal
polarization for the $J/\psi$ above 2 GeV/c. However, in the
CS frame, no conclusion can be drawn due to its limited
acceptance, besides the uncertainties in the
current data. Figure \ref{fig:charmonium2} (a) shows the transverse
momentum dependence of the $J/\psi$ polarization parameter in the HX
frame together with currently available theoretical models: COM
\cite{chung-2009}  and the s-channel cut CSM \cite{haberzettl:032006}. 

\begin{wrapfigure}{r}{0.49\textwidth}
  \begin{center}
\vspace{-30pt}
    {\includegraphics[width=0.49\textwidth]{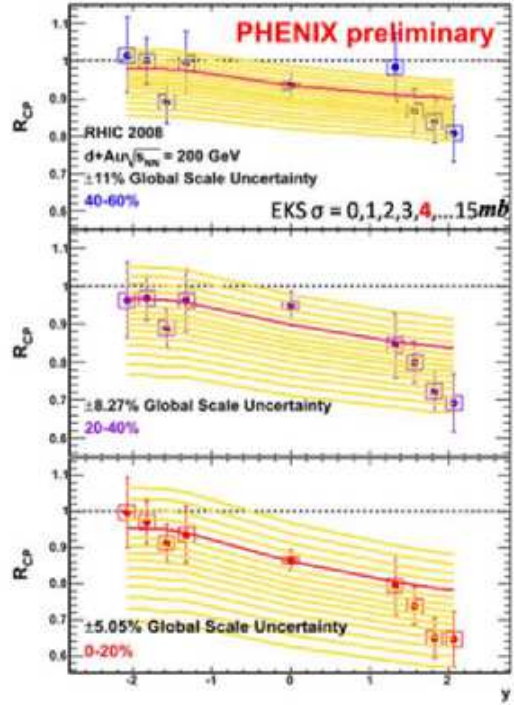}}
 \end{center}
\label{fig:dau}
\vspace{-20pt}
  \caption{Central to peripheral $ J/\psi$ production
    as a function of rapidity in three centrality ranges, together with the prediction assuming gluon distribuion modification
EKS98 \cite{1126-6708-2008-07-102} for different breakup cross
sections ($\sigma$, 0-15 mb).} 
\end{wrapfigure} 

\noindent Figure \ref{fig:charmonium2} (b) shows the transverse
momentum dependence of the $J/\psi$ polarization parameter in the GJ
frame.

PHENIX has recently presented preliminary results of dielectrons in the
$\Upsilon(1S+2S+3S)$ mass range of [8.5-11.5] GeV/$c^2$ in $p$+$p$ and
Au+Au collisions. Figure 8 shows the nuclear modification factor
$R_{AA}$ which was
evaluated with Poisson distribution and was found to be lower
than 0.64 in 90\% confidence level. This result may suggest that the $\Upsilon$ states 2S+3S are melted in the medium formed in Au+Au collisions. However, no final conclusion can be drawn before the knowledge of the cold nuclear matter effects in $d$+Au collisions and a better understanding of the continuum contribution in this mass range.

During the 2008 data taking period with $d$+Au collisions, PHENIX
recorded a factor of 30 higher statistics than the 2003 $d$+Au
Run. Preliminary studies of the $J/\psi$ yield ratio between central
and peripheral events ($R_{CP}$), as a function of rapidity, show that the current parton
modification functions and uniform breakup cross section in the
hadronic matter cannot describe the behavior of the $J/\psi$
suppression in $d$+Au collisions. Figure 9 shows the measured $R_{CP}$
in three centrality ranges, together with the prediction assuming gluon distribuion modification
EKS98 \cite{1126-6708-2008-07-102} for different breakup cross
sections ($\sigma$, 0-15 mb).

\section{Conclusions}

PHENIX has measured single electrons in both $p$+$p$ and Au+Au
collisions at midrapidity and single muons in $p$+$p$ collisions at forward
rapidity at $\sqrt{s_{NN}}$ = 200 GeV.  The ratio of the yield of electrons
from bottom to that from charm in $p$+$p$ collisions at the same value
of the c.m. energy provided the first measurement of the spectrum of
electrons from bottom at RHIC. All these results were presented in
the context of pQCD calculations. The electron-muon azimuthal
correlation result provided a clear measurement of heavy flavor
production in a rapidity range not yet studied. Quarkonia yields
in $p$+$p$ collisions included the $J/\psi$,  the $\Upsilon$, and the first $\psi^{\prime}$ 
differential cross sections at RHIC energies, besides the first
$J/\psi$ polarization measurement for two polarization frames,
providing additional insight into the understanding of the production mechanisms.

\section*{References}


\begin{thebibliography}{10}

\bibitem{Adcox2003469}
PHENIX Collaboration, K.~Adcox {\em et~al.}, {\em PHENIX detector overview},
\newblock Nuclear Instruments and Methods in Physics Research Section A:
  Accelerators, Spectrometers, Detectors and Associated Equipment {\bf 499},
  469  (2003),
\newblock The Relativistic Heavy Ion Collider Project: RHIC and its Detectors.

\bibitem{dion-2009}
A.~Dion, {\em Open Heavy Flavor at PHENIX},
\newblock arXiv.org:0907.4749  (2009).

\bibitem{adare:252002}
PHENIX Collaboration, A.~Adare {\em et~al.}, {\em Measurement of High-$p_T$
  Single Electrons from Heavy-Flavor Decays in p + p Collisions at $\sqrt{s}$ =
  200 GeV},
\newblock Phys. Rev. Lett. {\bf 97}, 252002 (2006).

\bibitem{adare:172301}
PHENIX Collaboration, A.~Adare {\em et~al.}, {\em Energy Loss and Flow of Heavy
  Quarks in Au + Au Collisions at $\sqrt{s_{NN}}$ = 200 GeV},
\newblock Phys. Rev. Lett. {\bf 98}, 172301 (2007).

\bibitem{PhysRevLett.95.122001}
M.~Cacciari, P.~Nason, and R.~Vogt, {\em QCD Predictions for Charm and Bottom
  Quark Production at RHIC},
\newblock Phys. Rev. Lett. {\bf 95}, 122001 (2005).

\bibitem{adler:092002}
PHENIX Collaboration, S.~S. Adler {\em et~al.}, {\em Measurement of single
  muons at forward rapidity in $p$+$p$ collisions at $\sqrt{s}$ = 200 GeV and
  implications for charm production},
\newblock Physical Review D (Particles and Fields) {\bf 76}, 092002 (2007).

\bibitem{0954-3899-35-10-104113}
D.~Hornback, {\em Measurements of heavy-quark production via single leptons at
  PHENIX},
\newblock Journal of Physics G: Nuclear and Particle Physics {\bf 35}, 104113
  (5pp) (2008).

\bibitem{adare:232301}
PHENIX Collaboration, A.~Adare {\em et~al.}, {\em $J/\psi$ Production versus
  Centrality, Transverse Momentum, and Rapidity in Au + Au Collisions at
  $\sqrt{s_{NN}}$ = 200 GeV},
\newblock Phys. Rev. Lett. {\bf 98}, 232301 (2007).

\bibitem{adare:082002}
PHENIX Collaboration, A.~Adare {\em et~al.}, {\em Measurement of Bottom Versus
  Charm as a Function of Transverse Momentum with Electron-Hadron Correlations
  in p + p Collisions at $\sqrt{s}$ = 200 GeV},
\newblock Phys. Rev. Lett. {\bf 103}, 082002 (2009).

\bibitem{engelmore-2009-830}
T.~Engelmore, {\em Heavy Flavor Production and Energy Loss with Two-Particle
  Correlations at PHENIX},
\newblock Nuclear Physics A {\bf 830}, 853c (2009).

\bibitem{PhysRevLett.93.171801}
F.~Cooper, M.~X. Liu, and G.~C. Nayak, {\em $J/\Psi{}$ Production in $p+p$
  Collisions at $\sqrt{s}$ =200 GeV at the BNL Relativistic Heavy Ion
  Collider},
\newblock Phys. Rev. Lett. {\bf 93}, 171801 (2004).

\bibitem{haberzettl:032006}
H.~Haberzettl and J.~P. Lansberg, {\em Possible Solution of the $J/\psi$
  Production Puzzle},
\newblock Phys. Rev. Lett. {\bf 100}, 032006 (2008).

\bibitem{dasilva-2009-830}
C.~L. {da Silva}, {\em Quarkonia measurement in p+p and d+Au collisions at
  $\sqrt{s}$ =200 GeV by PHENIX Detector},
\newblock Nuclear Physics A {\bf 830}, 227c (2009).

\bibitem{refId}
HERA-B Collaboration, I.~Abt {\em et~al.}, {\em A Measurement of the to
  $J/\psi$ to $\psi^{\prime}$ ratio in 920 proton-nucleus interactions},
\newblock Eur. Phys. J. C {\bf 49}, 545 (2007).

\bibitem{PhysRevLett.79.572}
CDF Collaboration, F.~Abe {\em et~al.}, {\em $J/\psi{}$ and $\psi{}(2S)$
  Production in $p+p\ifmmode\bar\else\textasciimacron\fi{}$ Collisions at
  $\sqrt{s}= $1.8 TeV},
\newblock Phys. Rev. Lett. {\bf 79}, 572 (1997).

\bibitem{collaboration-2009}
PHENIX Collaboration, A.~Adare {\em et~al.}, {\em Transverse momentum
  dependence of $J/\psi$ polarization at mid-rapidity in p+p collisions at
  $\sqrt{s}$ =200 GeV},
\newblock arXiv.org:0912.2082  (2009).

\bibitem{adare:232002}
PHENIX Collaboration, A.~Adare {\em et~al.}, {\em $J/\psi$ Production versus
  Transverse Momentum and Rapidity in p + p Collisions at $\sqrt{s}$ = 200
  GeV},
\newblock Phys. Rev. Lett. {\bf 98}, 232002 (2007).

\bibitem{1126-6708-2008-10-004}
P.~Faccioli, C.~Lourenco, J.~Seixas, and H.~K. Wohri, {\em Study of
  $\psi^{\prime}$ and $\chi_c$ decays as feed-down sources of $J/\psi$
  hadro-production},
\newblock Journal of High Energy Physics {\bf 2008}, 004 (2008).

\bibitem{PhysRevC.68.061902}
B.~L. Ioffe and D.~E. Kharzeev, {\em Quarkonium polarization in heavy ion
  collisions as a possible signature of the quark-gluon plasma},
\newblock Phys. Rev. C {\bf 68}, 061902 (2003).

\bibitem{chung-2009}
H.~S. Chung, S.~Kim, J.~Lee, and C.~Yu, {\em Polarization of prompt $J/\psi$ in
  proton-proton collisions at RHIC},
\newblock arXiv.org:0911.2113  (2009).

\bibitem{1126-6708-2008-07-102}
K.~J. Eskola, H.~Paukkunen, and C.~A. Salgado, {\em An improved global analysis
  of nuclear parton distribution functions including RHIC data},
\newblock Journal of High Energy Physics {\bf 2008}, 102 (2008).

\end{thebibliography}

\end{document}